\documentclass[nofootinbib,showkeys,twocolumn,aps]{revtex4}
\usepackage{epsf}

\begin{document}

\newcommand{\bgt}[1]{{\boldmath $#1$}}
\newcommand{\bgm}[1]{\mbox{\boldmath $#1$}}

\title{On the interaction of point charges\\
  in an arbitrary domain}

\author{Alexander Silbergleit} \email{gleit@relgyro.stanford.edu}

\author{Ilya Nemenman} \email{nemenman@research.nj.nec.com}
\altaffiliation{Current address: NEC Research Institute, 4
  Independence Way, Princeton, NJ 08540, USA}

\author{Ilya Mandel} \email{emandel@relgyro.stanford.edu}

\affiliation{Gravity Probe B, W.\ W.\ Hansen Experimental Physics
  Laboratory, \\Stanford University, Stanford, CA 94305-4085, USA}

\begin{abstract}
  We develop a systematic approach to calculating the electrostatic
  force between point charges in an arbitrary geometry with arbitrary
  boundary conditions. When the boundary is present, the simple
  expression for the force acting on a charge as ``the charge times
  the field it is placed in'' becomes ill-defined. However, this rule
  can be salvaged if the field in question is redefined to include all
  the terms that do not diverge at the charge position, in particular,
  those due to the charge itself. The proof requires handling the
  self-action energy divergence for point charges, which is
  accomplished by means of a geometrical regularization.
\end{abstract}



\date{\today}

\maketitle

\section{Introduction}

It is trivial to determine the force exerted by an external field
\footnote{In fact, the word ``external'' in any case means that the
  sources of the field, such as some boundaries, are far away from the
  charge.} on a point charge in an otherwise empty space: by
definition, ``the force is equal to the charge times the field it is
placed in''. In particular, if the field in question is created by
some other point charges, this rule, known by many from high school,
still holds.

However, the situation changes drastically when a set of point charges
creates the field inside an arbitrary domain with a boundary of some
physical origin (reflected in the appropriate boundary conditions).
Now the very notion of ``the field the charge is placed in'' becomes
not well defined. For example, a naive treatment of a single--charge
problem might lead one to an entirely wrong conclusion that, since
{\em all} the field in the problem is due to the charge itself (there
are no other sources!), ``the field it is placed in'' is zero, so
there is no force at all.

A slightly more sophisticated physicist would argue that only the part
of the field which diverges as $1/r^2$ near the charge is really
created by it, while the rest is due to the boundary conditions, which
represent mathematically the rearrangement of {\em other} physical
charges at the boundary.  Therefore it is precisely what remains after
subtracting the singular part that now gives ``the field the charge is
placed in''.  Unfortunately, such treatment leaves one in a somewhat
awkward position of, first, calculating potentials and fields
rigorously, and then lowering the plank and using hand-waving
arguments to derive forces from them. It is also not clear whether the
guess about which part of the total field contributes to the force is
always valid.

Thus it seems appealing to show that the physical arguments can be
backed by an accurate mathematical proof demonstrating that the {\em
  adjusted} rule, ``the force is equal to the charge times the part of
the field that does not diverge at the charge's location'', is either
universal or limited by certain conditions. To do this, one should
turn to the most fundamental energy conservation argument which gives
the force as the negative gradient of the energy in the charge's
position. This approach also does not turn out to be straightforward,
since the energy is infinite in the presence of point charges due to
their self--action.

Perhaps because of these difficulties, as well as of a misleading
apparent simplicity of the problem, our literature search, which
encompassed, in particular,
Refs.~\onlinecite{zomm,tamm,strat,llf,llc,smy,jack,feynman,cro,moo}
and many other books on the subject, revealed no ready result (except
for one small piece in Ref.~\onlinecite{smy} which we discuss in
Sec.~\ref{discussion}).  So we give a careful derivation of the
general expression for the force on point charges in this paper. It
consists of a regularization of the problem, calculation of the force
from the (regularized and finite) energy, and then taking the singular
limit.  The result agrees with one's intuitive expectations.

\section{Electrostatics problem with volume point charges: 
  potential and energy}

Consider an arbitrary 3-dimensional domain $D$ with the perfectly
conducting boundary $S$ and some $N$ point electrical charges inside.
The electrical potential, $\psi({\bf r})$, in this case is determined
by the following Dirichlet boundary value problem (we use SI units
throughout the paper):
\begin{eqnarray}
  \Delta \psi 
  &=&
  -\frac{1}{\varepsilon_0}\sum_{i=1}^N 
  q_i\,\delta\left({\bf r} - {\bf r}_i\right),\quad {\bf r},
  \,\,{\bf r}_i\in D;
  \label{pde}
  \\
  \psi\,\biggl|_{S} 
  &=& 
  0 \,.
  \label{boundary}
\end{eqnarray}
Here ${\bf r}=x{\bf e}_x+ y{\bf e}_y+z{\bf e}_z$ is the vector radius
of a point, and ${\bf r}_i=x_i{\bf e}_x+ y_i{\bf e}_y+z_i{\bf e}_z$
specifies the $i$th charge position, with ${\bf e}_\alpha,\,\,
\alpha=x,y,z$, being the unit vectors in the direction of the
corresponding Cartesian axes.

By superposition principle, the potential $\psi({\bf r})$ is just the
sum of the potentials induced by each charge separately,
\begin{eqnarray} 
  \psi({\bf r})&=&
  \kappa \sum_{j=1}^Nq_jG({\bf r},{\bf r}_j)
  \label{psi_Green}
  \\
  &\equiv&
  \kappa\sum_{j=1}^Nq_j\left[\frac{1}{|{\bf r} - 
      {\bf r}_j|}+G_R({\bf r},{\bf r}_j)\right],
  \label{psi_sing_R}
\end{eqnarray}
where $\kappa=1/4\pi \varepsilon_0$, and $G_R({\bf r},{\bf r}_j)$ is
the regular part of the Green's function $G({\bf r},{\bf r}_j)$ of the
corresponding boundary value problem [set $q_j=1,\,\,
q_i=0,\,\,i\not=j$ in Eq.~(\ref{pde})].  Both functions are, of
course, symmetric in their arguments,
\begin{equation} \label{Gsymm}
  G({\bf r},{\bf r}_j)=G({\bf r}_j,{\bf r}),
  \qquad G_R({\bf r},{\bf r}_j)=G_R({\bf r}_j,{\bf r}).
\end{equation}
Furthermore, we can rewrite Eq.~(\ref{psi_sing_R}) splitting the
potential in a sum of its singular and regular parts,
\begin{eqnarray}
  \psi({\bf r}) &=& \kappa\sum_{j=1}^N 
  \frac{q_j}{|{\bf r} 
    - {\bf r}_j|} + \psi_R({\bf r}),
  \label{psiparts}\quad
  \\
  \psi_R({\bf r}) 
  &\equiv&
  \kappa
  \sum_{j=1}^Nq_jG_R({\bf r},{\bf r}_j)\;,
  \label{psiR}
\end{eqnarray}
where $\psi_R({\bf r})$ is a regular function satisfying the Laplace
equation everywhere in $D$ [by continuity, this holds also at any
regular point \footnote{We allow for the boundary singularities, such
  as sharp edges and spikes, provided that the Meixner type finite
  energy condition~\cite{mili} is satisfied near them; in particular,
  the domain $D$ can be infinite.} of the boundary $S$, although this
is irrelevant to our discussion].  Note that both the potential,
$\psi$, and its regular part, $\psi_R$, depend actually on the
positions of the charges ${\bf r}_i$ as well as on the observation
point ${\bf r}$, which is reflected in the full notation,
\begin{eqnarray}
  \psi({\bf r}) &\equiv& \psi({\bf r},\,{\bf r}_1,
  \dots,{\bf r}_i,\dots,{\bf r}_N),
  \label{psifull}\quad
  \\
  \psi_R({\bf r}) 
  &\equiv&
  \psi_R({\bf r},\,
  {\bf r}_1,\dots,{\bf r}_i,\dots,{\bf r}_N).
  \label{psiRfull}
\end{eqnarray}

We assume that the potential is known and are interested in finding
the force ${\bf F}^i$ acting on the charge $q_i$. From the energy
conservation for the considered problem, the force is given by
(cf.~Ref.~\onlinecite{smy}) \footnote{It is important to understand
  that Eq.~(\ref{force}) is a {\em definition} of a mathematical
  object that we would like to correspond to the physical force. If
  there are no external fields [Eq.~(\ref{boundary}) ensures this in
  our discussion], it will turn out that the conclusions derived from
  Eq.~(\ref{force}) are physically meaningful and validate the
  definition.}
\begin{equation} \label{force}
  {\bf F}^i=
  -\frac{\partial}{\partial{\bf r_i}} 
  W_D\,, \;\;\; \frac{\partial}{\partial{\bf r_i}} =
  \frac{\partial}{\partial x_i}{\bf e}_x+ 
  \frac{\partial}{\partial y_i}{\bf e}_y
  +\frac{\partial}{\partial z_i}{\bf e}_z\; ,
\end{equation}
where $W_D$ is the energy of the field in the volume $D$,
\begin{equation} 
  \label{wd}
  W_D=\frac{\varepsilon_0}{2}\,\int_D  (\bgm{\nabla} \psi)^2\,dV.
\end{equation}
Note that we alternatively write $\bgm{\nabla}$ or
$\partial/\partial{\bf r}$ for the gradient, whatever seems proper in
a particular expression.

The problem is, however, that the above integral obviously diverges
due to self-interaction of the point charges (the energy of a single
point charge is infinite).  We are going to show that even though the
energy for a given point charge distribution is infinite, the {\it
  difference} between its two values corresponding to two different
but close charge configurations {\it is finite for any charge
  configuration and boundary shape}, and the {\it force is also
  finite} due to that, in accordance with common intuition.

\section{Regularized energy and the force on the charges}
\label{energy_force}

\begin{figure}[bp]
  \centerline{\epsfxsize=0.67\hsize\epsffile{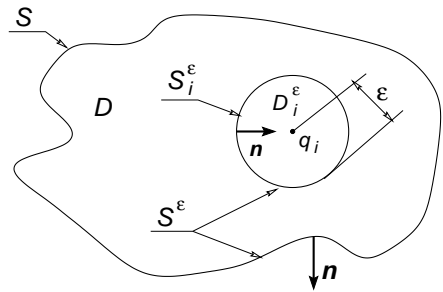}}
  \caption{Volumes, surfaces, and normal directions involved.}
  \label{fig:domain}
\end{figure}

We surround each volume charge $q_i$ by a small sphere
$S_i^{\epsilon}$ of radius $\epsilon$; we write $D_i^{\epsilon}$ for
the ball inside it.  We define $D^\epsilon$ as $D$ without all domains
$D_i^{\epsilon}$, and $S^\epsilon$ as a union of ${S}$ and all
spherical surfaces $S_i^{\epsilon}$ (see Fig.~\ref{fig:domain}).  In
effect, $S^\epsilon$ is the boundary of $D^\epsilon$.

Using Eq.~(\ref{force}), we may now define the force acting on the
charge $q_i$ as
\begin{equation} 
  {\bf F}^i= \lim_{\epsilon \to 0}{\bf F}^i_\epsilon
  =-\lim_{\epsilon \to 0} 
  \frac{\partial}{\partial{\bf r_i}} W_D^{\epsilon},
  \label{forcelimit}
\end{equation}
where $W_D^{\epsilon}$ is the regularized energy, that is, the energy
of the field in $D^\epsilon$, which is finite. It is important to note
the order of operations in Eq.~(\ref{forcelimit}): first take the
gradient of the regularized energy in the charge position, then take
the (singular) limit. In principle, we also have to show that the
final result does not depend on the regularization chosen, but this
task is not easy. We will return to it briefly later in this paper.

In view of Eq.~(\ref{wd}) and the fact that the total potential given
by Eq.~(\ref{psi_Green}) or Eq.~(\ref{psiparts}) is regular in
$D^\epsilon$, the regularized energy is
\begin{eqnarray}
  W_D^{\epsilon}&\equiv&\,\frac{\varepsilon_0}{2} \int_{D^\epsilon}
  (\bgm{\nabla} \psi)^2\,d V
  \\  
  &=& \frac{\varepsilon_0}{2}\int_{S^\epsilon}
  \psi \frac{\partial\psi}{\partial {n}}\, d{A} -
  \frac{\varepsilon_0}{2}\int_{D^\epsilon} \psi\Delta\psi\,dV;
\end{eqnarray}
$n$ is the direction of the {\it outward} normal to $S^\epsilon$ (and
thus the {\it inward} normal to the spheres $S_i^{\epsilon}$).  For an
infinite domain $D$ it is assumed here that the potential and its
gradient drop at infinity fast enough to make the contribution of
integrating over the sphere of a large radius vanishing in the limit,
which assumption has to be verified in each particular case.

Since $\psi$ is harmonic everywhere in $D^\epsilon$, the volume
integral on the right of the previous equality vanishes; the remaining
surface one is represented as
\begin{equation}
  W_D^{\epsilon}= \frac{\varepsilon_0}{2}\sum_{k=1}^N
  \int_{S_k^{\epsilon}} \psi \frac{\partial\psi}{\partial {n}} \,d {A} +
  \frac{\varepsilon_0}{2}\int_{S} \psi \frac{\partial\psi}{\partial {n}}
  \, d {A}\label{WeDfull}
\end{equation}
and then, because of the boundary condition, Eq.~(\ref{boundary}), as
\begin{equation} \label{WeD}
  W_D^{\epsilon}=  \frac{\varepsilon_0}{2}\sum_{k=1}^N 
  \int_{S_k^{\epsilon}} \psi \frac{\partial\psi}{\partial {n}}\,d {A} .
\end{equation}

We are ultimately interested in the limit $\epsilon \to 0$, so we need
to calculate only the quantities which do not vanish in this limit.
The area of integration in each term of the above sum is
$O(\epsilon^2)$, therefore we need to keep track of the integrands
that grow at least quadratically in $\epsilon^{-1}$. Bearing this in
mind and using Eq.~(\ref{psiparts}) for the potential, we can write
the surface integral in Eq.~(\ref{WeD}) as
\begin{widetext}
  \begin{equation}\label{intterms}
    \int_{S_k^{\epsilon}} \psi \frac{\partial\psi}{\partial {n}} \,d{A}
    = \int_{S_k^{\epsilon}} \frac{\kappa \, q_k}{|{\bf r}
      - {\bf r}_k|} \,\frac{\partial}{\partial {n}}
    \left(\frac{\kappa \, q_k}{|{\bf r} - {\bf
          r}_k|}\right)\,d{ A} +
    \int_{S_k^{\epsilon}} \left[ \psi_R ({\bf r}) +\sum_{j=1, j \neq k}^N
      \frac{\kappa \,q_j}{|{\bf r} - {\bf r}_j|} \right] \,
    \frac{\partial}{\partial {n}}\left(\frac{\kappa \, q_k}
      {|{\bf r} - {\bf r}_k|}\right)\, 
    d{A} +O(\epsilon) \; .
  \end{equation}
\end{widetext}

The first term in the above expression is, in fact, a regularized
self--energy of the $k$-th charge, $W_{k,\, self}^{\epsilon}$. Doing
an elementary integration, we immediately find that
\begin{equation} \label{Wself}
  W_{k,\, self}^{\epsilon}=\frac{\varepsilon_0}{2}
  \,  \kappa^2 \,
  \frac{4 \pi q_k^2}{\epsilon}=
  \frac{\kappa \, q_k^2}{2 \epsilon} \, .
\end{equation}
The only feature of the regularized self--energy given by
Eq.~(\ref{Wself}) important for our derivation is that {\it it does
  not depend on the position of the charge $q_k$, i.\ e., on the
  vector radius ${\bf r}_k$}.

The second term of the r.\ h.\ s.\ of Eq.~(\ref{intterms}) can also be
simplified if one notices that both $\psi_R$ and $1/| {\bf r} - {\bf
  r}_j|$, $j\neq k$, are regular on $S_k^{\epsilon}$ and in
$D_k^{\epsilon}$. Therefore their change within the small surface
$S_k^{\epsilon}$ is of order $\epsilon$.  Thus Eq.~(\ref{intterms})
may be rewritten as
\begin{widetext}
  \begin{eqnarray}\label{moreint}
    \frac{\varepsilon_0}{2} \int_{S_k^{\epsilon}} \psi
    \frac{\partial\psi}{\partial {n}} \,d{A}&=& {W_{k, self}^{\epsilon}} 
    + \frac{\varepsilon_0}{2} \left[ \psi_R({\bf r_k}) +
      \sum_{j=1,\,j \neq k}^N \frac{\kappa \, q_j}{|{\bf
          r}_k - {\bf r}_j|} \right] 
    \int_{S_k^{\epsilon}} \frac{\partial}{\partial {n}}
    \left(\frac{\kappa\,q_k}{|{\bf r} - {\bf r}_k|} \right)\,d{A}
    +O(\epsilon)
    \nonumber
    \\
    &=&
    W_{k, self}^{\epsilon}+\frac{q_k}{2} \left[ \psi_R ({\bf r_k})+
      \sum_{j=1,\, j \neq k}^N \frac{\kappa \, q_j}{|{\bf
          r}_k - {\bf r}_j|} \right] +O(\epsilon),
  \end{eqnarray}
  and the integration here yielding the factor $4\pi$ is again an
  elementary one. This asymptotic equality may be differentiated in
  ${\bf r}_i$ with the same estimate of the remainder term.
  
  Introducing now the last expression into the Eq.~(\ref{WeD}), we
  obtain:
  \begin{eqnarray} 
    W_D^{\epsilon} &=& \sum_{k=1}^N W_{k,\, self}^{\epsilon} \; +\;
    \frac{\kappa}{2}
    \sum_{k=1}^N \sum_{j=1,\, j\neq k}^N 
    \frac{q_j \,q_k}{|{\bf r}_j-{\bf r}_k|}\; 
    +\;
    \frac{1}{2}\sum_{k=1}^N q_k\psi_R({\bf r}_k)\; + \; O(\epsilon)\; .
    \label{WeDfinal}
  \end{eqnarray}
  
  Equation (\ref{WeDfinal}), in its turn, is inserted in the
  Eq.~(\ref{forcelimit}) for the force; as shown, the self--energies
  do not depend on the charge positions, hence, although diverging in
  the limit $\epsilon \to 0$, {\it they do not contribute to the
    force}. The rest is pretty straightforward, except one has to be
  careful when differentiating the last term on the right of
  Eq.~(\ref{WeDfinal}) with $k=i$: as seen from Eq.~(\ref{psifull}),
  in this case ${\bf r}_i$ stands for {\it two} (and not one!)
  arguments of $\psi_R$, namely, $\psi_R({\bf r_i}) \equiv \psi_R({\bf
    r}_i,{\bf r}_1,\dots,{\bf r}_i,\dots,{\bf r}_N)$, and {\it both of
    them} have to be differentiated.  Bearing this in mind, the
  expression for the force finally becomes:
  \begin{eqnarray}
    {\bf F}^i&=& - \kappa q_i\sum_{k=1, k \neq i}^N    
    \frac{\partial}{\partial{\bf r_i}}\frac{q_k
      }{|{\bf r}_i - {\bf r}_k|}\;
    -\;\frac{1}{2} \left[ \sum_{k=1}^{N} q_k\frac{\partial}{\partial{\bf
          r_i}}\psi_R({\bf r})\biggl|_{\bf r=\bf r_k}+
      q_i\frac{\partial}{\partial{\bf r}}\psi_R({\bf r})\biggl|_{{\bf
          r}={\bf r}_i} \right]
    \nonumber
    \\
    &=&
    \label{14}\label{ffirst}
    \kappa q_i\sum_{k=1, k \neq i}^N   q_k
    \frac{{\bf r}_i -{\bf r}_k}{|{\bf r}_i - {\bf r}_k|^3} \;
    -\;\frac{1}{2}
    \left[
      \sum_{k=1}^{N} q_k\frac{\partial}{\partial{\bf r_i}}
      \psi_R({\bf r})\biggl|_{\bf r=\bf r_k}+
      q_i \frac{\partial}{\partial{\bf r}}\psi_R({\bf r})
      \biggl|_{{\bf r}={\bf r}_i}
    \right] \; .
  \end{eqnarray}

  This is the general result for the electrostatics which can be
  transformed further in some nice way. Indeed, the direct substitution
  of the expression for $\psi_R$ from the Eq.~(\ref{psiparts}) in the
  the Eq.~(\ref{ffirst}) provides the force in the form
  \begin{eqnarray}
    {\bf F}^i &=& 
    -\kappa q_i
    \sum_{k=1, k \neq i}^N \frac{\partial}{\partial{\bf r_i}}\frac{q_k}{|{\bf r}_i - {\bf
        r}_k|}\;
    -\;
    \frac{\kappa q_i}{2} 
    \left[ 
      \sum_{k=1}^{N}
      q_k\frac{\partial}{\partial{\bf r_i}}G_R({\bf r},{\bf r}_i)
      \biggl|_{\bf r=\bf r_k}+ \sum_{j=1}^{N}q_j
      \frac{\partial}{\partial{\bf r}}G_R({\bf r},{\bf r}_j)
      \biggl|_{{\bf r}={\bf r}_i} \right]
    \nonumber
    \\
    &=& 
    \label{15} \label{fsymm}                             
    -\kappa q_i\left[
      \sum_{k=1, k \neq i}^N 
      \frac{\partial}{\partial{\bf r_i}}\frac{q_k}{|{\bf r}_i - {\bf
          r}_k|}\;
      +\;
      \sum_{k=1}^{N} q_k \frac{\partial}{\partial{\bf r}}
      G_R({\bf r},{\bf r}_k)\biggl|_{{\bf r}={\bf r}_i}
    \right],
  \end{eqnarray}
  and we have used the symmetry property of Eq.~(\ref{Gsymm}) to obtain
  the second equality here. To make the result even more physically
  transparent, we rewrite Eq.~(\ref{fsymm}), in its turn, in the
  following way:
  \begin{equation}  \label{ffinal}
    {\bf F}^i= -
    \kappa q_i
    \bgm{\nabla}\left\{
      \sum_{k=1}^{N}q_k\left[ \frac{1}{|{\bf r} - {\bf r}_k|} +
        G_R({\bf r},{\bf r}_k)\right]-\frac{q_i}{|{\bf r} - {\bf r}_i|}
    \right\}\Biggl|_{{\bf r}={\bf r}_i}=
    -q_i\bgm{\nabla}\left[
      \psi({\bf r})- 
      \frac{\kappa \, q_i}{
        |{\bf r} - {\bf r}_i|}
    \right]\Biggl|_{{\bf r}={\bf r}_i}\;.
  \end{equation}
\end{widetext}
Note that the last expression, indeed, coincides with our intuitive
guess for the form of the force.

\section{Discussion}
\label{discussion}

Our first remark on the expressions for the force in
Eqs.~(\ref{ffirst})--(\ref{ffinal}) is that for the charges in a free
space (volume $D$ is the whole space, no boundaries present)
apparently $G_R({\bf r},{\bf r}_k)\equiv0,\,\psi_R\equiv0$, and the
classical Coulomb formula for the force is restored.

Next, Eq.~(\ref{ffinal}) shows that the rule ``the force is the charge
times the field it is placed in'' does work {\it if one counts the
  regular part of the field produced by the charge in question as a
  part of the ``field the charge is placed in''.} It also makes up to
some ``minimal principle'', namely: to get the right answer for the
force, one should throw out of the field only the part which otherwise
makes the result infinite, {\it and nothing beyond that}.  As we
mentioned in the Introduction, this result is supported by physical
intuition. It becomes even more transparent if one notes that the
singular part of the field thrown out is radial, and the radial field
produces no force.

A contribution of the regular part of the field created by a charge to
the force acting on it is especially important in the case of a single
charge, as one may see from the simplest example of a charge near a
conducting plane. It is exactly the regular part of the field produced
by the charge in question (equal to the field of the image charge)
that gives the whole answer when no other charges are present.

Finally, an important question is how {\it robust} our regularization
of the problem is, i.\ e., whether the result for the force does not
change if one uses a different regularization. There are two
significant points demonstrating such robustness.

The first one is concerned with the {\it geometrical} regularization
that we used. If one chooses domain $D_k^\epsilon$ around $q_k$ to be
not a ball but some differently shaped volume bounded by smooth
surface $S_k^\epsilon$ (``topological ball''), then it is not
difficult to see that all the terms of order $O(\epsilon)$ in
Eq.~(\ref{WeDfinal}) for the regularized energy remain unchanged, and
hence our result for the force is still true. The demonstration goes
exactly in the same way as above, only the computation of the integral
over the surface $S_k^\epsilon$ in Eq.~(\ref{moreint}) requires a
well--known result from potential theory (cf.~Ref.~\onlinecite{kell}).
As for the first integral on the right of Eq.~(\ref{intterms}), which
defines the self--energy $W_{k,\, self}^{\epsilon}$, its explicit
expression is not even needed, and its only relevant property, namely,
its independence of ${\bf r_k}$, is obvious.

An alternative way of regularization, so widely used by the classics
during the whole ``pre-Dirac delta-function'' era, is the {\it
  physical} regularization, when the point charge $q_k$ is replaced,
within the small volume $D_k^\epsilon$, with some smooth charge
distribution of the density $\rho_k^\epsilon({\bf r})$ and the same
total charge $q_k$, and $\epsilon$ is taken to zero in the answer.
From the technical point of view, this approach proves to be more
complicated in this particular case, but it leads again to the same
$O(\epsilon)$ terms in Eq.~(\ref{WeDfinal}) for the regularized
energy. The key point here is to start with the following expression
for the regularized energy,
\begin{equation}
  W_D^{\epsilon}\,\equiv\,
  \frac{1}{2}\,\int_{D} \rho^\epsilon\,\psi\,dV\,=\,
\frac{1}{2}\,\sum_{k=1}^N\,\int_{D_k^\epsilon} \rho_k^\epsilon\,\psi\,dV\;,
\end{equation}
and then, instead of Eq.~(\ref{psi_Green}), split the potential into a
sum of volume potentials of $\rho_k^\epsilon({\bf r})$ over
$D_k^\epsilon$ (which becomes singular in the limit), and a regular
addition $\psi_R^\epsilon({\bf r})$. In particular, this
regularization is used by Smythe in Sec.~3.08 of Ref.~\onlinecite{smy}
for calculating the force on a single point charge in a domain with
the zero potential at the boundary. The derivation there is at the
`physical level of accuracy', and the answer is not brought down to
its physically most relevant form of Eq.~(\ref{ffinal}).  Moreover,
the final answer there [r.\ h.\ s.\ of Eq.~(2) in that Section] is,
unfortunately, formally diverging, because of the inappropriate use of
the notation for the total potential in place where its regular part
should be.

Finally, we want to end our discussion by noticing that the
electrostatic problem we just solved, as well as its generalizations
(see Sec.~\ref{generalizations}), involve only volume charges.  On the
other hand, magnetostatic problems that deal, for example, with
magnetic fluxes trapped in superconducting media
(cf.~Ref.~\onlinecite{tink}) give rise to surface charges. Analysis of
these is of extreme importance for today's experimental physics
\cite{ns}. No easy solution for the force between surface charges
should be anticipated since the details of the boundary shape, such as
its curvature, are expected to play a role; the interaction of such
surface charges will be discussed elsewhere.

\section{Generalization: other boundary conditions}
\label{generalizations}

We can now generalize our result for other conditions at the boundary.
A modest but potentially useful generalization is to the case of
electrodes, when an arbitrary distribution of the potential, $V({\bf
  r})$, and not just a zero, is specified at the boundary:
\begin{equation}\label{17}
  \psi\,\biggl|_{S} =V({\bf r}),\qquad {\bf r}\in S \;.
\end{equation}
Let us split the potential in two,
\begin{equation}\label{split}
  \psi=\psi^{(1)}+\psi^{(2)}\, ,
\end{equation}
of which the first is caused by point charges without any voltage
applied to the boundary, and the second is entirely due to the
boundary voltage. Therefore $\psi^{(1)}$ satisfies the boundary value
problem of Eqs.~(\ref{pde}) and (\ref{boundary}),
\begin{eqnarray}
  \Delta \psi^{(1)} &=&-\frac{1}{\varepsilon_0}\sum_{i=1}^N 
  q_i\,\delta\left({\bf r} - {\bf r}_i\right),\quad {\bf r},
  \,{\bf r}_i\in D;
  \label{pde1}
  \\
  \psi^{(1)}\,\biggl|_{S} &=& 0 \; .
  \label{boundary1}
\end{eqnarray}

According to what is proved above, the force on a charge from the
field specified by the potential $\psi^{(1)}$ is given according to
Eq.~(\ref{ffinal}),
\begin{equation}
  {\bf F}^i_{(1)}= -q_i\bgm{\nabla}\left[
    \psi^{(1)}({\bf r})- 
    \frac{\kappa \, q_i}{ |{\bf r} - {\bf r}_i|}
  \right]\Biggl|_{{\bf r}={\bf r}_i}\;.
  \label{forceV1}
\end{equation}
On the other hand, potential $\psi^{(2)}$, satisfying
\begin{equation}
  \Delta \psi^{(2)} =0,\quad{\bf r}\in D;
  \qquad \psi^{(2)}\,\biggl|_{S}
  =V({\bf r})\; , \label{psi2}
\end{equation}
describes the field {\it external} to all the point charges, since it
does not depend on them and their positions. Therefore the force
exerted by this field is
\begin{equation} 
  {\bf F}^i_{(2)}= -q_i\bgm{\nabla}
  \psi^{(2)}({\bf r})\Biggl|_{{\bf r}={\bf r}_i}\;.
  \label{forceV2}
\end{equation}
Using now the superposition principle, we add these two forces to
reinstate the result of Eq.~(\ref{ffinal}) in the considered case:
\begin{equation}\label{ffinalV}
  {\bf F}^i= {\bf F}^i_{(1)}\;+\;{\bf F}^i_{(2)}\;=\;                            
  -q_i\bgm{\nabla}
  \left[
    \psi({\bf r})-\frac{\kappa \, q_i}{|{\bf r} - {\bf r}_i|}
  \right]
  \Biggl|_{{\bf r}={\bf r}_i}\;.
\end{equation}

The mixed boundary conditions
\begin{equation} 
  \psi\,\biggl|_{S_1} =V({\bf r}),\qquad
  \varepsilon_0\,\frac{\partial\psi}
  {\partial{n}}\,\biggl|_{S_2} =\sigma({\bf r})\; ,
\end{equation} 
where the surfaces $S_1,S_2$ are non-intersecting ($S_1\cap
S_2=\emptyset$) and comprise the whole boundary ($S_1\cup S_2=S$), and
$V({\bf r}),\,\sigma({\bf r})$ are given functions, lead to the same
standard result for the force [Eq.~(\ref{ffinal})] without any new
technical difficulties. Indeed, we split the total potential in two as
in Eq.~(\ref{split}) and require that
\begin{eqnarray}
  \Delta \psi^{(1)} &=&-\frac{1}{\varepsilon_0}\sum_{i=1}^N 
  q_i\,\delta\left({\bf r} - {\bf r}_i\right),\quad {\bf r},
  \,\,{\bf r}_i\in D;
  \\
  \psi^{(1)}\,\biggl|_{S_1} &=&0,\qquad \frac{\partial\psi^{(1)}}{\partial
    {n}}\,\biggl|_{S_2} =0 \; ,
  \label{psimix1}
\end{eqnarray}
and
\begin{eqnarray}
  \Delta \psi^{(2)} &=&0,
  \quad{\bf r}\in D;
  \\
  \psi^{(2)}\,\biggl|_{S_1} &=&V({\bf r}),\qquad 
  \varepsilon_0\,\frac{\partial\psi^{(2)}}{\partial
    {n}}\,\biggl|_{S_2} =\sigma({\bf r}) \; .
  \label{psimix2}
\end{eqnarray}
The derivation of the force from $\psi^{(1)}$ goes exactly as in
Sec.~\ref{energy_force} and leads to Eq.~(\ref{forceV1}). The external
to the charges field from $\psi^{(2)}$ produces the force of
Eq.~(\ref{forceV2}), so by superposition the total force is again as
in Eq.~(\ref{ffinal}) [or Eq.~(\ref{ffinalV})].

The appropriate split of the potential in two parts
[Eq.~(\ref{split})] is a little bit trickier for the Neumann
boundary condition,
\begin{equation} 
\varepsilon_0\,\frac{\partial\psi}{\partial
  {n}}\,\biggl|_{S}=\sigma({\bf r}),\;\;
\int_S\,\sigma({\bf r})\,dA\;+\;{Q}=0\;, \;\; Q\equiv\sum_{j=1}^N  q_j.
\end{equation} 
Namely, the solvability criterion (the total charge must be zero)
makes us, when splitting the potential, to add and subtract another
charge $Q$ (equal to the sum of the point charges $q_i$) at some point
${\bf r_*}$ of the domain $D$, to obtain
\begin{eqnarray}
  \Delta \psi^{(1)} &=&-\frac{1}{\varepsilon_0}
  \left[
    \sum_{i=1}^N 
    q_i\,\delta\left({\bf r} - {\bf r}_i\right)-Q\,
    \delta\left({\bf r} - {\bf r}_*\right)
  \right],
  \nonumber
  \\
  &&{\bf r},
  \,\,{\bf r}_i,\,\,{\bf r}_*\in D;\qquad
  \\
  \frac{\partial\psi^{(1)}}{\partial
  {n}}\,\biggl|_{S} &=&0 \; ,
  \label{psiN1}
\end{eqnarray}
as well as
\begin{eqnarray}
  \Delta \psi^{(2)} &=&-\frac{Q}{\varepsilon_0}\,
  \delta\left({\bf r} - {\bf r}_*\right),
  \quad {\bf r},\,\,{\bf r}_*\in D;
  \\
  \varepsilon_0\,\frac{\partial\psi^{(2)}}{\partial
  {n}}\,\biggl|_{S} &=&\sigma({\bf r}) \; ,
  \label{psiN2}
\end{eqnarray}
with both problems solvable. Again, the derivation of the force from
$\psi^{(1)}$ satisfying the homogeneous boundary condition goes
exactly as before and leads to Eq.~(\ref{forceV1}), the external to
the charges field $\psi^{(2)}$ exerts the force given in
Eq.~(\ref{forceV2}), and by superposition the result of
Eq.~(\ref{ffinal}) holds. The problem itself, though, is not too
realistic, except for the case of an insulated boundary, $\sigma({\bf
  r})\equiv0$.

\begin{acknowledgments}
  This work was supported by NASA grant NAS 8-39225 to Gravity Probe
  B.  In addition, I.\ N.\ was partially supported by NEC Research
  Institute.  The authors are grateful to R.\ V.\ Wagoner and V.\ S.\ 
  Mandel for valuable remarks, and to GP-B Theory Group for a fruitful
  discussion.
\end{acknowledgments}

\end{document}